# Ontology-based question answering over corporate structured data

Sergey Gorshkov[1] [0000-0002-0821-8050], Constantin Kondratiev[1],
Roman Shebalov[1]

[1] TriniData LLC, 40-21 Mashinnaya str., 620089, Ekaterinburg, Russia
`serge@trinidata.ru, kondratiev@trinidata.ru,`
`shebalov@trinidata.ru`

**Abstract.** Ontology-based approach to the Natural Language Understanding (NLU) processing allows to improve questions answering quality in dialogue systems. We describe our NLU engine architecture and evaluate its implementation. The engine transforms user's input into the SPARQL SELECT, ASK or INSERT query to the knowledge graph provided by the ontology-based data virtualization platform. The transformation is based on the lexical level of the knowledge graph built according to the Ontolex ontology.

The described approach can be applied for graph data population tasks and to the question answering systems implementation, including chat bots. We describe the dialogue engine for a chat bot which can keep the conversation context and ask clarifying questions, simulating some aspects of the human logical thinking. Our approach uses graph-based algorithms to avoid gathering datasets, required in the neural nets-based approaches, and provide better explainability of our models. Using question answering engine in conjunction with data virtualization layer over the corporate data sources allows extracting facts from the structured data to be used in conversation.

## 1    Introduction

The corporate automated system users expect them to be "intellectual" enough to give precise answers to their questions. It implies that a system must give meaningful answers in the dialogue with a user and ask clarifying questions as a human would do. To achieve that, the system must deal with the structured representation of each question and answer of a dialogue, as well as with a structured data required to give an answer. The knowledge graphs (KGs) are one of the most popular ways to represent the complex structured information.

KGs can be populated with information by data mining from text sources or structured data sets – for example, the business application databases in the corporate environment. Some aspects of the corporate KGs assembly are considered in [Noy, 2019].

Our task is to develop the complex solution (framework) for the corporate use which will provide:

1) Conceptual model construction for representing the users' domain knowledge. We do not have a task of automated ontology assembly or enrichment because in the corporate projects the ontology is predominantly composed manually.
2) Disparate corporate data sets into a virtual Knowledge Graph.
3) The natural language tools for user interaction with KG.

The domain ontology composition and the virtual KG assembly are out of scope of this paper (see: [Gorshkov, 2021]). We will focus on the methods of user's natural language question transformation into the SPARQL query to the knowledge graph, and the dialogue system implementation.

## 2      Task definition

We consider the KG question-answering and the task of facts extraction from the natural language text as the task of the representing the text meaning in the form of the graph restricted by a conceptual model. In this paper we focus on the description of the dialogue system which transforms a user's question into a query to the graph. The functional requirements to this system are:

1) It must find the class or property which instances would be an answer to the user's question and query the KG to find the appropriate entities matching criteria defined in the question.
2) The system's answer should be precise, not probabilistic. The system must be able to prove that the found objects are the answers to the user's query, including visualization of the relations chain that led to these objects.
3) If there is no unambiguous answer to the user's query, the system should ask clarifying questions.
4) The system should answer user's clarifying questions, which may be asked if the answer is not comprehensive for the user. It means that the system shall keep the conversation context, i.e., the objects and relations mentioned in the previous questions and answers.

These requirements cannot be fulfilled today using the neural networks only. The researchers [Thorne, 2021] note that the contemporary NL processing models cannot scale to non-trivial databases nor answer set-based and aggregation queries. Any neural network model output is probabilistic by its nature and cannot be provided with the proof of correctness. We believe, following [Weikum, 2021], that machine knowledge and machine learning complement and strengthen each other. We aimed to combine optimally the strong points of both approaches, machine learning and logical inference, when designing the text processing pipeline.

The neural networks are effective in knowledge graph query answering (KGQA) when dealing with the big graphs containing ambiguities, such as DBPedia. The big datasets are required to train such models. This is often impossible when dealing with corporate tasks in the specific and narrow domains. Due to these factors, we have set a goal of creating the explainable and fully controlled tool.

Let us describe a domain which we will use as a field for evaluating our solution. Consider the sample industrial enterprise which is composed of the functional units and sites. Each unit and site have a person or organizational unit which is responsible for some aspect of its safety (fire, industrial, etc.). All this information is gathered into a KG, which also contains data on the telemetry sensors and the parameters they measure. We have created a compact ontology for this domain, which offers diversity of the relations between objects and the playground for making queries involving 3-4 related graph vertices. In the real use case such a system will include a much extensive set of facts on the various aspects of the enterprise activity. The facts will be gathered into KG from the variety of data sources, such as corporate applications' databases, the organizational and administrative documents, and will be consolidated by a data virtualization platform. The resulting graph should be available with SPARQL interface.

## 3       Related works

We have used some well-known technologies in our NL processing pipeline. Named entities recognition techniques developed over long time [Shen, 2015]. POS tagging is considered in [Mikic, 2009; Wu, 2020; Huang, 2015; Le, 2018; Piccinno, 2014]. Since our ideas are based on moving from syntactical relations to semantical relations (see: [Melchuk, 1999; Gerd, 2005; Kolshansky, 1980; Banarescu, 2013; Fensel, 2003]), we need a morphological and syntactical analysis of sentences and coreference clusters (chains) finding. Morpho-syntactical properties of tokens, analysis of syntactic structure of a phrase, syntactic relationships discovery between words are considered in [Jurafsky, 2008]. Co-reference resolution techniques review is given in [Zheng, 2011].

Pre-trained language models such as BERT [Kuratov, 2019] can be used to determine the context-depending word meaning. Transformer models allow vectorization of the word sequences, which can be used to interpret their meanings [Kalyan, 2021]. The ability to retrieve an entity from a Knowledge Base given a textual input is a key component for several applications (see: [Ferrucci, 2012; Slawski, 2015; Yang, 2018]).

We have used Ontolex[1] ontology, developed by W3C Ontology-Lexica Community Group[2], to formalize lexical model. This ontology was published in 2016[3] and it is well documented. Its key features are described in its developer's publications, for example [Cimiano, 2011], [McCrae, 2017]. Ontolex is often used in the computational linguistics tasks, for examples see: [Declerck, 2019, Abgaz, 2020]. It is also used in the interdisciplinary projects, particularly it was used in the European Commission PMKI project (Public Multilingual Knowledge Infrastructure) in 2018[4].

---

## 4 Implementation

Our NL processing pipeline includes several principal steps: the preliminary text analysis (including morphological, syntactical analysis and coreference resolution), search for the ontology entities mentions, adding predicates between the mentions, uncertainties resolution, adding hidden nodes and choice of winners, individual objects resolution, and the result assembly. At the last step we compose and execute SPARQL query and produce the human-readable answer if a query represents a user's question. We will describe each of these steps in details below.

All these steps are performed successively. During the execution we build a graph containing edges and vertices of various types, and at each step we enrich it. We have developed the framework which includes a text processing pipeline. Each step is implemented as a separate module and can be replaced if needed, for example, with the ML model. It also includes a dialogue engine for chat bot scenarios definition, as well as some supplementary tools for results visualization, text processing, graph construction and working with a SPARQL endpoint.

### 4.1 Preliminary text analysis

At this stage the text is split into sentences, then sentences are split into tokens. Then the morphological and syntactical analysis is performed. We use a machine learning model that uses RuBERT [Kuratov, 2019]. For the most of the preliminary text processing tasks we use DeepPavlov[5] library.

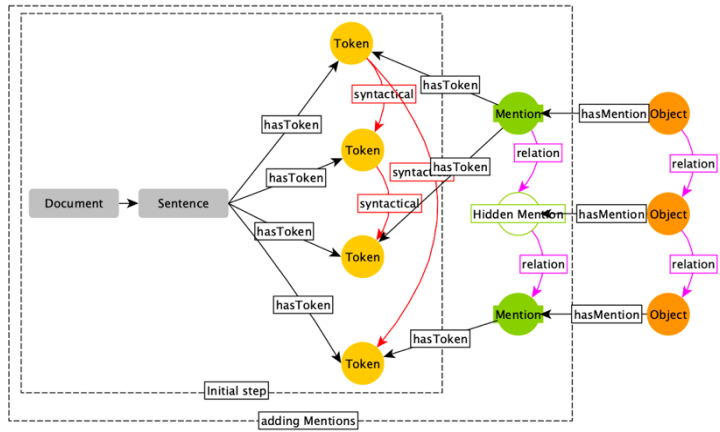

Fig. 1. The document graph. Initial step - at the end of preliminary text analysis. Adding mentions - the search for ontology entities mentions and uncertainties resolution.

We complement each token with the morphological properties and connect tokens by the edges according to the Universal Dependencies specification[6]. At this stage we also perform coreference resolution using DeepPavlov library. The document graph D, which is composed as a result of this stage, is shown in Figure 1 (initial step).

---

[5] https://deeppavlov.ai

[6] https://universaldependencies.org

## 4.2 Search for ontology entities mentions

Our lexical model is based on the Ontolex ontology which includes classes to denote lexical objects and relate them to the domain elements. In Figure 2 we show the diagram of the core part of the ontology we use to express such relations (word meanings) and unite them in the semantic fields.

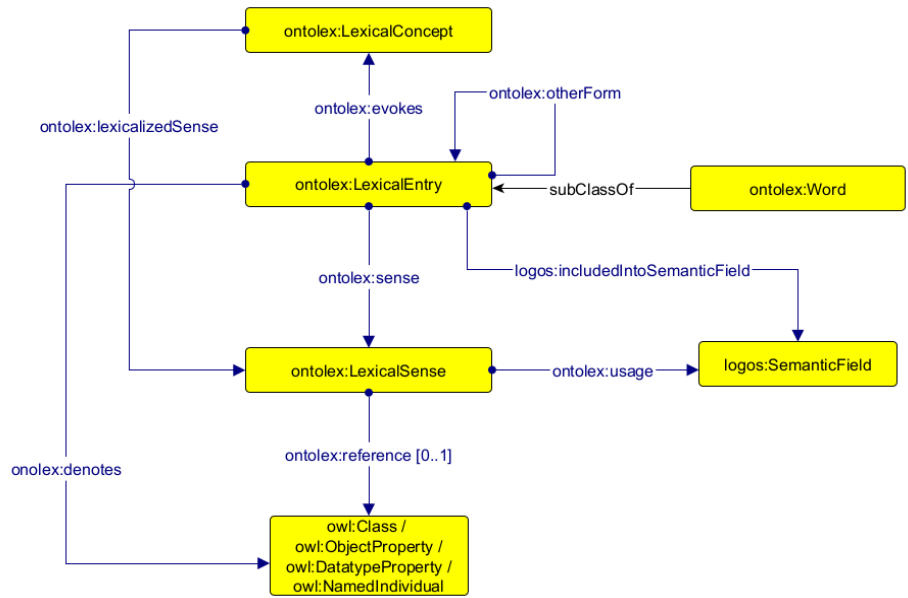

Figure 2. The core part of the linguistic ontology used in our pipeline

The "Word" class takes the central place in this model. Its instances are the words used in the texts. Each word can be related with some concept or lexical sense (LexicalSense class). Lexical sense is the lexical meaning of the word, its denotation in some context. We use lexical senses to narrow the word meaning, if necessary, for some domain. The lexical senses usage allows to get rid of connotations and emotional, evaluative components which a word can possess in particular situations.

The lexical model includes also the SemanticField class we introduced to extend Ontolex. Its individuals allow to distinguish ambiguous words semantics. Each semantic field consists of the words which share some core meaning component. The word inclusion into some SemanticField allows to narrow its meaning in this context. This feature is used to handle lexical uncertainties including those based on polysemy, homonymy, homography, etc.

The task of ontology entities mentions search can be described as the task of concepts search in the text. Each concept can consist of one or more tokens. For the mor-phologically rich languages, such as Russian, the word order in the text is almost random. It means that the tokens representing one concept can be spaced to the different parts of the phrase.

We break up a text into the syntactically related words sequences and bind each sequence with the Mention. The Mention for us is a possible ontology entity reference in the text. Each token can present in the names of several ontology entities, and it can reference these entities in a phrase with certain probability.

We have implemented several matchers for this step. A matcher is a program code class with the specific interface, which searches for mentions in the text. The matchers can search for words, for the named entities (we use a standard NER list including Person, Location etc., using DeepPavlov library instruments). A matcher can use string templates: we use the regular expressions and the Yargy[7] library, as well as search of graph entities by rdfs:label predicate. Each mention is assigned with the weight, which reflects probability that some group of words is related with some element of an ontology. The structure of the document graph D at the end of this step is presented in Figure 1 (adding mentions).

### 4.3 Adding predicates between mentions

The practical approaches to the natural language sentences analysis are based on the various theoretical assumptions and hypotheses. Particularly, the whole possibility of the formal sentence decomposition aimed to determine its lexical and grammatical parts makes us turn to the question of semantic dominants that can be probably found as a result. If we assume that syntactically bound elements represent certain semantic dependencies, this question can be classified as a context analysis problem. We intentionally omit the problem of the "context-free grammar" [Chomsky, 1956] and so-called "context-independent languages" [Jurafsky, 2008] and are using the "meaning representation" approaches [Melchuk, 1999; Gerd, 2005; Kolshansky, 1980; Banarescu, 2013; Fensel, 2003].

As each mention contains the text tokens, and the tokens have syntactical relations, we consider the subgraph $K(T, S)$, where $T \subset V(D)$ are all the vertices representing Tokens, and $S \subset E(D)$ is a subset of all the edges representing syntactical relations. The vertices $t \in T$ are related by the patches of the length N, and the tokens are the parts of the Mentions: $h_i = \{t_i, m_i\}$, $h_k = \{t_k, m_k\}$, where $m \in M$, $M \subset V(D)$ is a subset of all the vertices representing Mentions, $h_i \in H$, $H \subset E(D)$ is a subset of all the edges representing hasTokens relation. If there is an edge ($e=\{m_i, m_k\}$, $e \in E(G)$) in the domain ontology graph, it is a reason for adding a semantic relation (predicate) $r=\{m_i, m_k\}$ into the document graph D.

### 4.4 Uncertainties resolution, adding hidden nodes and winners choice

At this stage each Token of the text may be referenced by several Mentions, and there may exist several semantic vertices between Mentions. Consider subgraphs $A(M,$

---

R) and B(T∪M, H), where T ⊂ V(D) are all the vertices of Token type, M ⊂ V(D) are the Mentions of the document graph, R ⊂ E(D) are the edges of Relation type, and H ⊂ E(D) is a subset of vertices of hasToken type. The task of selecting the most probable Mentions can be considered as the "max flow min cost" task in the B graph. It can be intuitively described as the choice of the Mention (M), when the cost flow is minimized (cost is a value reciprocal of the vertex weight). We use NetworkX[8] library for this task, which has a suitable implementation of this algorithm. The weight of each vertex M is produced from the weight of the Mention found by matcher and the vertex degree in the A subgraph (min cost).

The text may contain ellipses (an omission of a sentence element, which is reconstructed by a reader with the help of context), then the structure of (D) graph derived from the text will not match the domain ontology subgraph (G). In this case the number of weakly connected components of D graph will be more than one. At this stage we add new "hidden" nodes, which refer to the ontology entities not mentioned explicitly in the phrase, to the D graph to minimize the number of weakly connected components.

We perform also some extra transformations and choose from a number of hypotheses the one most probable subgraph of ontology graph, which contains "hidden" nodes and filtered Mentions. The graph structure at the end of this step is shown in Figure 1.

### 4.5    Individual objects resolution

Consider the following text: "In the first tank of the gas liquefaction unit… in the second tank… in the third tank…". In this text the `PlantUnit` class is mentioned twice, and the Tank class is mentioned three times. As humans we can assume that the same individual object of `PlantUnit` class is mentioned, as well as three different individuals of the `Tank` class. At this stage of the text processing pipeline, we have to resolve these mentions and find the particular individual objects. Each `Tank` is related with `PlantUnit` by the `isPartOf` predicate. We shall filter individuals using these predicates and consider predicates cardinalities (`owl:maxCardinality`):

```
:isPartOf rdf:type owl:ObjectProperty ;
  rdfs:domain :Tank ;
  rdfs:range :PlantUnit .
:hasNumber rdf:type owl:DatatypeProperty ;
  rdfs:domain :Tank ;
  rdfs:range xsd:int ;
  owl:maxCardinality 1 .
```

With ontology in mind we can derive that there are  three numbered individuals of Tank class (`hasNumber owl:maxCardinality 1`) and one `PlantUnit`. We use also coreference clusters for the individuals search.

Then we copy predicates from the Mentions to the Objects. The graph structure at the end of this step is shown in Figure 1.

---

### 4.6 Dialogue management

The above figures show the structure of the graph containing Tokens, Mentions, Objects, etc. We have implemented a utility for this graph visualization. The real graph produced from the question "Ivanov's phone" is shown in Figure 7.

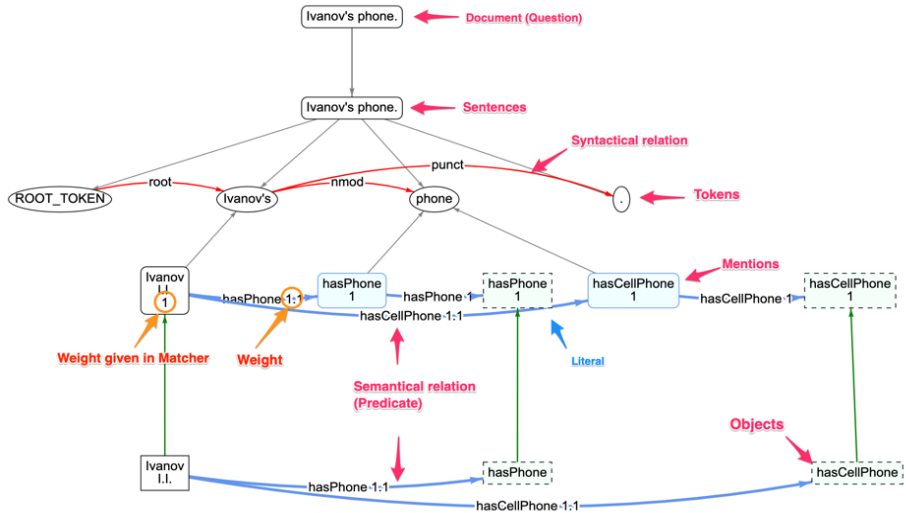

Figure 7. The graph produced from the question "Ivanov's phone"

This illustrates how we represent a text in the form of a graph. This is suitable for answering a single question. We can also perform the tasks of facts extraction from a text and of the simple one-hop search (one answer for one given question). But we have a goal of creating an intellectual assistant which can interact with a user. We have to resolve the next tasks to achieve the following: dialogue forking, context support, user question augmentation and clarification.

### 4.7 Context support

We support the approaches used in the classic chat-bots, implemented as the final state machine. At each moment of time this dialogue agent exists in some state. The states transition rules are set, so the dialogue scenario is defined imperatively. We have implemented a special module for the states switching and rules processing.

### 4.8 User question augmentation and clarification. Evaluation

We have implemented a demo interface to evaluate our method. Chat bot answers the questions, posing the clarifying questions if necessary, and answer user's clarifying questions. The questions can cover any objects of the domain ontology described in the Task definition section. The data from the customer's corporate systems, consolidated in the KG using ontology-based data virtualization platform, are used to provide answers. Consider the dialogue started with a user's question: "Who is responsible for the fire safety of the gas liquefaction units?". After passing our pipeline, the following SPARQL request is formed:

```
SELECT * WHERE {
    ?psr rdf:type base:PersonalSafetyResponsibility .
    ?psr base:hasSafetyAspect base:FireSafety .
    ?psr base:hasPerson ?person .
    ?org rdf:type org:OrganizationalUnit .
    ?org    base:operates    base:1d8dc36f-909d-4711-a1cd-
1ae74b305e9d .
    ?person rdf:type foaf:Person .
    ?person org:memberOf ?org }
```

In this query, `base:1d8dc36f-909d-4711-a1cd-1ae74b305e9d` is an identifier of the individual object representing gas liquefaction unit found in the ontology. Note the ?org vertice of `OrganizationalUnit` class – it was added by the engine automatically as the shortest way between the `foaf:Person` class individual and the gas liquefaction unit. The human-readable representation of this query result:

Petrov Petr
    class: Person
    Is an employee of the unit: Gas liquefaction units.

Let a user ask the next question: "Which is his phone?". To answer this question, we have to infer that a user asks for Petrov's phone. We concatenate both phrases and build the graph representing them. During the coreference chains analysis we find a pronominal anaphor: the pronoun "his" refers to the token "who" from the first phrase. This token, in turn, is a Mention of the `foaf:Person` class, and the word "phone" is a Mention of `hasPhone` predicate. At this stage we have two graphs representing the first and second questions. Two edges of these graphs are related by `hasCoreference` edge. We can complement the previous query with the following patterns:

```
    ?person rdf:type foaf:Person .
    ?person org:memberOf ?org .
    ?person base:hasPhone ?phone .
```

If the pronominal or nominal anaphors are not found, the algorithm binds two graphs by looping all the connections possible according to the domain ontology, minimizing path lengths and the number of weakly connected components of the united graph.

In some cases, the algorithm will ask clarifying questions if it cannot produce reasonable query from a question or obtain a reasonable result. Consider the user's question "Smith's phone". At the mentions search step we discover using NER model that "Smith" is the human's last name. This leads to the following query:

```
    ?s rdf:type foaf:Person;
    foaf:familyName "Smith"
```

If this query does not give an answer, or returns too many objects, the algorithm asks a clarifying question to the user. The user will make a more specific query to obtain a reasonable result.

# 5    Conclusions

In the quest of the most effective combination of the machine learning and KG tools we have developed the architecture of the natural language understanding pipeline. The algorithms of establishing links between tokens of the recognized text and the domain ontology play the key role in it. The lexical ontology layer, describing words usage to denote domain concepts, is necessary to make them work. It allows semantic ambiguities resolution considering semantic fields in which the words are included.

The practical value of the developed architecture significantly increases if the KG queries it generates to answer user questions are processed by the data virtualization platform which can access the huge arrays of the data disparate in the corporate storages. It opens the way to create a dialogue system which allows user to discover previously hidden, implicit, or virtually unavailable information from these storages, and involves it in the business processes including decision making support. This way of KG usage can be described as the valuable component of the true Knowledge Management System of an enterprise.

The further work on our NL solution improvement includes quantitative question answering implementation (object counting, searching for maximal/minimal values, summation, and other aggregation methods), work with the date and numeric intervals, temporal relations recognition. These functions are valuable in the corporate data processing.